\documentclass[conference]{IEEEtran}
\usepackage{blindtext, graphicx, amsmath, amsfonts, amssymb}

\usepackage{cite}

\usepackage{url}

\usepackage[table,svgnames]{xcolor}

\usepackage{multirow}

\usepackage{relsize}

\newcommand\blfootnote[1]{%
  \begingroup
  \renewcommand\thefootnote{}\footnote{#1}%
  \addtocounter{footnote}{-1}%
  \endgroup
}

\DeclareMathOperator*{\argmax}{arg\,max}

\begin{document}


\title{Positioning of High-speed Trains using \\ 5G New Radio Synchronization Signals}


\author{\IEEEauthorblockN{Jukka Talvitie\IEEEauthorrefmark{1}, Toni Levanen\IEEEauthorrefmark{1}, Mike Koivisto\IEEEauthorrefmark{1}, Kari Pajukoski \IEEEauthorrefmark{2}, Markku Renfors \IEEEauthorrefmark{1}, Mikko Valkama \IEEEauthorrefmark{1}\\}

\IEEEauthorblockA{\IEEEauthorrefmark{1}Laboratory of Electronics and Communications Engineering, Tampere University of Technology, Finland \\}
\IEEEauthorblockA{\IEEEauthorrefmark{2}Nokia Bell Labs, Finland \\}
Email: jukka.talvitie@tut.fi}




\maketitle

\begin{abstract}

We study positioning of high-speed trains in 5G new radio (NR) networks by utilizing specific NR synchronization signals. The studies are based on simulations with 3GPP-specified radio channel models including path loss, shadowing and fast fading effects. The considered positioning approach exploits measurement of Time-Of-Arrival (TOA) and Angle-Of-Departure (AOD), which are estimated from beamformed NR synchronization signals. Based on the given measurements and the assumed train movement model, the train position is tracked by using an Extended Kalman Filter (EKF), which is able to handle the non-linear relationship between the TOA and AOD measurements, and the estimated train position parameters. It is shown that in the considered scenario the TOA measurements are able to achieve better accuracy compared to the AOD measurements. However, as shown by the results, the best tracking performance is achieved, when both of the measurements are considered. In this case, a very high, sub-meter, tracking accuracy can be achieved for most (\textgreater 75\%) of the tracking time, thus achieving the positioning accuracy requirements envisioned for the 5G NR. The pursued high-accuracy and high-availability positioning technology is considered to be in a key role in several envisioned HST use cases, such as mission-critical autonomous train systems.

\end{abstract}


%

\blfootnote{This work was partially supported by the Finnish Funding Agency for Technology and Innovation (Tekes) and Nokia Bell Labs, under the projects ”Wireless for Verticals (WIVE)”, ”Phoenix+” and ”5G Radio Systems Research”, and by the Academy of Finland (under the projects 276378, 288670, and 304147).}

\section{Introduction}
\label{sec:introduction}

The forthcoming 5G new radio (NR) developments open new carrier frequencies for mobile communications which together with the wide millimeter-wave channels allow to design a dedicated high-speed train (HST) multi-gigabit wireless network \cite{TR_38913_scenario}. Moreover, the new 5G networks introduce a unique opportunity for high-accuracy and high-availability positioning \cite{5G_mike,5G_Cui}. Depending on the considered use case for the position information, a submeter positioning accuracy is pursued with a use-case-specific availability requirement. Whereas with non-critical use cases it is sufficient to achieve the submeter accuracy only on average rate, with safety-critical uses cases, such as autonomous trains \cite{autonomous_trains}, the minimum availability requirement can be higher than 99.9999~\%. In the considered scenario, the train acts as a relay, aggregating the traffic generated by the HST passengers and relaying it in downlink and uplink. Because we can assume a more sophisticated hardware solution in the train, we can assume a considerable amount of antenna elements, good processing power, and sufficient amount of memory in the device. In addition to low latency and high throughput communications, the 5G NR provides an excellent radio interface to accurately track the train location either within the train side or the network side.

The need for additional positioning technology is relevant, as Global Navigation Satellite Systems (GNSS) alone cannot guarantee consistent and accurate positioning performance \cite{ETCS_odometry_req,multisensor_train_pos}. Especially in urban environment, with high-rise buildings and urban canyons, GNSS performance is decreased due to strong multipath propagation and reduced satellite visibility. In addition, GNSS is known to be vulnerable to various malicious attacks including jamming and spoofing \cite{GPS_vulnerabilities}. Conventionally, the dependability on the GNSS has been mitigated by equipping trains with multiple positioning sensors, such as inertial navigation sensor, tachometer, and Doppler radar \cite{multisensor_train_pos}. Unfortunately, these types of in-train sensors are rather sensitive to cumulative error propagation, and are not able to provide a stand-alone positioning solution. However, recent advances in 5G technology, including specified requirements for sub-meter positioning accuracy \cite{TR_22862_pos_req}, have made network-based positioning an attractive option to complement future train positioning systems.

\begin{figure*}[ht]
    \centering
    \includegraphics[width=\textwidth]{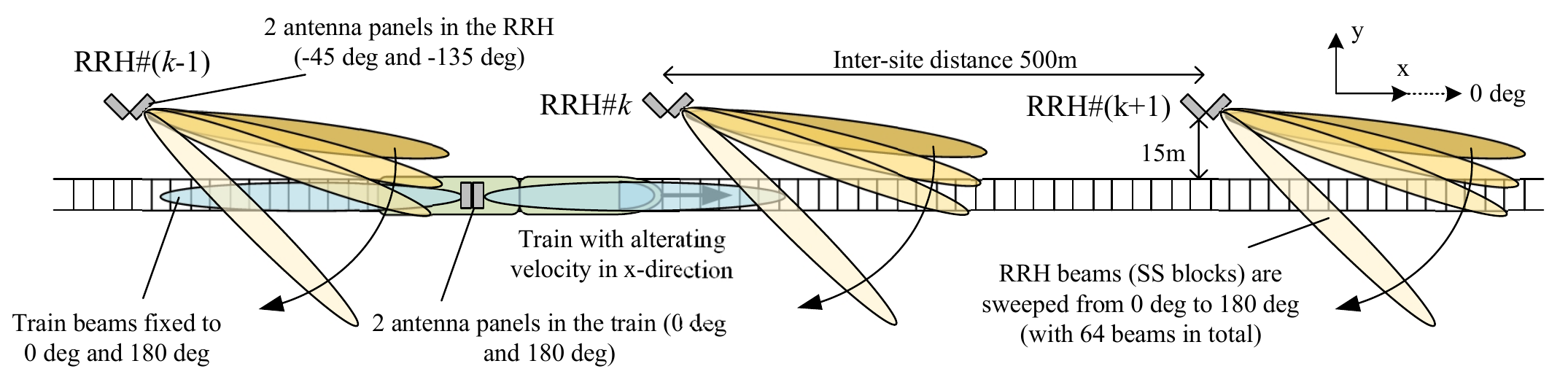}
    \caption{Illustration of the train radio transmission and positioning scenario. The RRHs are located beside the track and they send their SS block burst sets in 5ms intervals. The beamformer sweeps the track area by transmitting each SS block in the SS burst set in separate direction while the antenna beams in the train are assumed fixed along the track.}
    \label{fig:system}
\end{figure*}

Wireless network-based positioning has been widely studied in literature, see, for example, \cite{wireless_pos_survey}. Positioning with current 4G Long Term Evolution (LTE) networks is considered, for example, in \cite{LTE_pos1} and \cite{LTE_pos2}, and positioning with forthcoming 5G networks, for example, in \cite{5G_mike,5G_mike_cars,5G_Cui}. However, due to absence of detailed descriptions of 5G signal and frame structures, the 5G positioning studies have been limited to consider only assumed signal models. On contrary to this, in this paper we consider positioning of the HST based on the 5G NR downlink Synchronization Signals (SS), recently specified by the 3GPP \cite{TS_38211_signal}. In the HST network context, these signals are transmitted by Remote Radio Heads (RRH) controlled by a 5G NR Baseband Unit (BBU). The synchronization signal consists of Primary Synchronization Signal (PSS), Secondary Synchronization Signal (SSS), and Physical Broadcast Channel (PBCH), from which the PSS and SSS signals have very good built-in correlation properties allowing accurate Time-Of-Arrival (TOA) and Angle-Of-Departure (AOD) measurements.

By using the SS, we consider positioning on the train side, where TOA and AOD measurements, together with Extended Kalman Filter (EKF), are used to track the position, velocity, and acceleration of the train. Here, the main use case for high accuracy positioning is autonomous operation of the train. Even at very high velocities, e.g., 400-500 km/h, we show that the studied SS-based train tracking accuracy can be pushed close to 1~m accuracy. This allows the automated train to accurately control its location, velocity, and acceleration to minimize travel times and energy consumption and to maximize passenger safety and quality of service for the train passengers. Having an accurate localization service based on the 5G NR communication service, it also allows to reduce the maintenance costs of dedicated train positioning solutions applied, such as track balises.

The rest of this paper is organized as follows. In Section \ref{sec:system_model}, we describe the considered radio transmission and train positioning scenario, and define the assumed radio propagation models and 5G NR synchronization signal structures. Then, in Section \ref{sec:TOA_AOD_estimation}, we develop TOA and AOD estimators utilizing the received 5G NR synchronization signals. After describing the EKF-based train tracking approach in Section \ref{sec:position_estimation}, the performance of the proposed approach is evaluated and analyzed in Section \ref{sec:results}. Finally, conclusions are drawn in Section \ref{sec:conclusions}.


\section{System Model}
\label{sec:system_model}

The assumed HST network scenario, illustrated in Fig. \ref{fig:system}, is based on the 3GPP-specified high-speed scenario with 30~GHz carrier deployment given in \cite{TR_38913_scenario}. In this scenario, RRHs with beamforming capabilities are distributed along the track in order to serve the trains. Next, in this section, we go through the most fundamental system model assumptions, in terms of radio propagation, transmitted downlink signal structures, and assumed antenna models in the RRHs and in the train.

\subsection{Large scale radio propagation and antenna models}
\label{sec:large_scale_propagation}

At a position $\mathbf{p}(t)=\left[x(t),y(t)\right]^{\text{T}}$ the average received signal power (dBm) of a signal transmitted from a single RRH located at $\mathbf{p}_{0}(t)=\left[x_{0}(t),y_{0}(t)\right]^{\text{T}}$ is defined as
\begin{multline} \label{eq:received_power}
	P_{\text{R}}(\mathbf{p}(t)) = \\
    P_{\text{T}} - L(d(t)) + S(\mathbf{p}(t)) + G_{\text{T}}(\theta_{\text{T}}(t)) + G_{\text{\text{R}}}(\theta_{\text{R}}(t))
\end{multline}
where $P_{\text{T}}$ is the transmit power (dBm), $L(d(t))$ is the path loss (PL) (dB) given as a function of the propagation distance $d(t)=\lVert \mathbf{p}(t)-\mathbf{p}_{0}(t) \rVert$, and $S(\mathbf{p}(t))$ is the shadowing function (dB). Moreover, $G_{\text{T}}(\theta_{\text{T}}(t))$ and $G_{\text{R}}(\theta_{\text{R}}(t))$ are the beamforming gains (dB) given as a function of AOD $\theta_{\text{T}}(t)$ and AOA $\theta_{\text{R}}(t)$, where AOD and AOA are defined with respect to the antenna orientations of the transmitter and receiver.

We consider the urban Micro (uMI) PL model with Line-Of-Sight (LOS) path and a Gaussian distributed shadowing function with appropriate spatial correlation, as described in \cite[Section 7.4]{TR_38900_channel}. In addition, we define the beamforming gains $G_{\text{T}}(\theta_{\text{T}})$ and $G_{\text{R}}(\theta_{\text{R}})$ by using a beam pattern obtained from a Uniform Linear Array (ULA) with $M_{\text{T}}$ and $M_{\text{R}}$ horizontal antenna elements for the transmitter and receiver, respectively. Furthermore, we assume that the RRHs are able to do beam-steering while the train has fixed beams along the track.

\subsection{Fast fading and the received signal model}
\label{sec:fast_fading}

The fast fading channel model is based on the multipath channel specified by the Tapped Delay Line D (TDL-D) model given in \cite[Table 7.7.2-4]{TR_38900_channel} with a RMS delay spread of 20~ns. The maximum Doppler shift is given as $\Delta f_{\text{MAX}}=|v|/\lambda_{\text{c}}$, where $v$ is the velocity of the receiver and $\lambda_{\text{c}}$ is the wavelength of the carrier. Now, the received signal from a single RRH at a time instant $t$ can be written as
\begin{equation} \label{eq:received_signal}
\begin{gathered}
    z(t) = \tilde{z}(t) + v(t) \text{, where} \\
    \tilde{z}(t) = \gamma(t) \int_{-\infty}^{\infty} b(\xi)h(t-\xi,\xi) d\xi
\end{gathered}
\end{equation}
and where $b(t)$ is the transmitted signal, $h(t-\xi,\xi)$ is the channel impulse response to an impulse sent at time $\xi$, and $v(t)$ is additive white Gaussian noise. Moreover, we normalize the power of $b(t)$ and the expected total power of all multipath components of $h(t-\xi,\xi)$ to unity. Consequently, the scaling function $\gamma(t) = 10^{P_{\text{R}}(\mathbf{p}(t))/10}$ incorporates the effect of the large scale propagation model, including the beamforming gains, as described in (\ref{eq:received_power}). The format of $h(t-\xi,\xi)$ is considered to be unaltered throughout separate beams.

\subsection{Transmitted signal structure}
\label{sec:signal_structure}

The transmitted signal is based on a Cyclic-Prefix Orthogonal Frequency Division Multiplexing (CP-OFDM) waveform described in 3GPP 5G NR specifications \cite{TS_38211_signal}. As recommended for the high-speed scenarios, we have decided to use the subcarrier spacing of $\Delta f=240$~kHz for enhanced Doppler resistance. In this paper, we assume a total of 50 scheduled Physical Resource Blocks (PRBs), which results in the passband width of $50 \cdot 12 \cdot \Delta f = 144\text{MHz}$. Furthermore, we choose the Fast Fourier Transform (FFT) size as $N_{\text{FFT}}=1024$ (with 600 active subcarriers), and hence, the basic physical layer processing rate as $F_{\text{s}}=245.76$~MHz. The length of one subframe is 1ms, including 16 time slots, in which each time slot consists of 14 OFDM symbols. As specified in \cite{TS_38211_signal}, the first symbol and the middle symbol in the subframe have an extended CP length of 200 samples, whereas all other OFDM symbols have a normal CP length of 72 samples. 

\begin{figure}[t]
    \centering
    \includegraphics[width=\columnwidth]{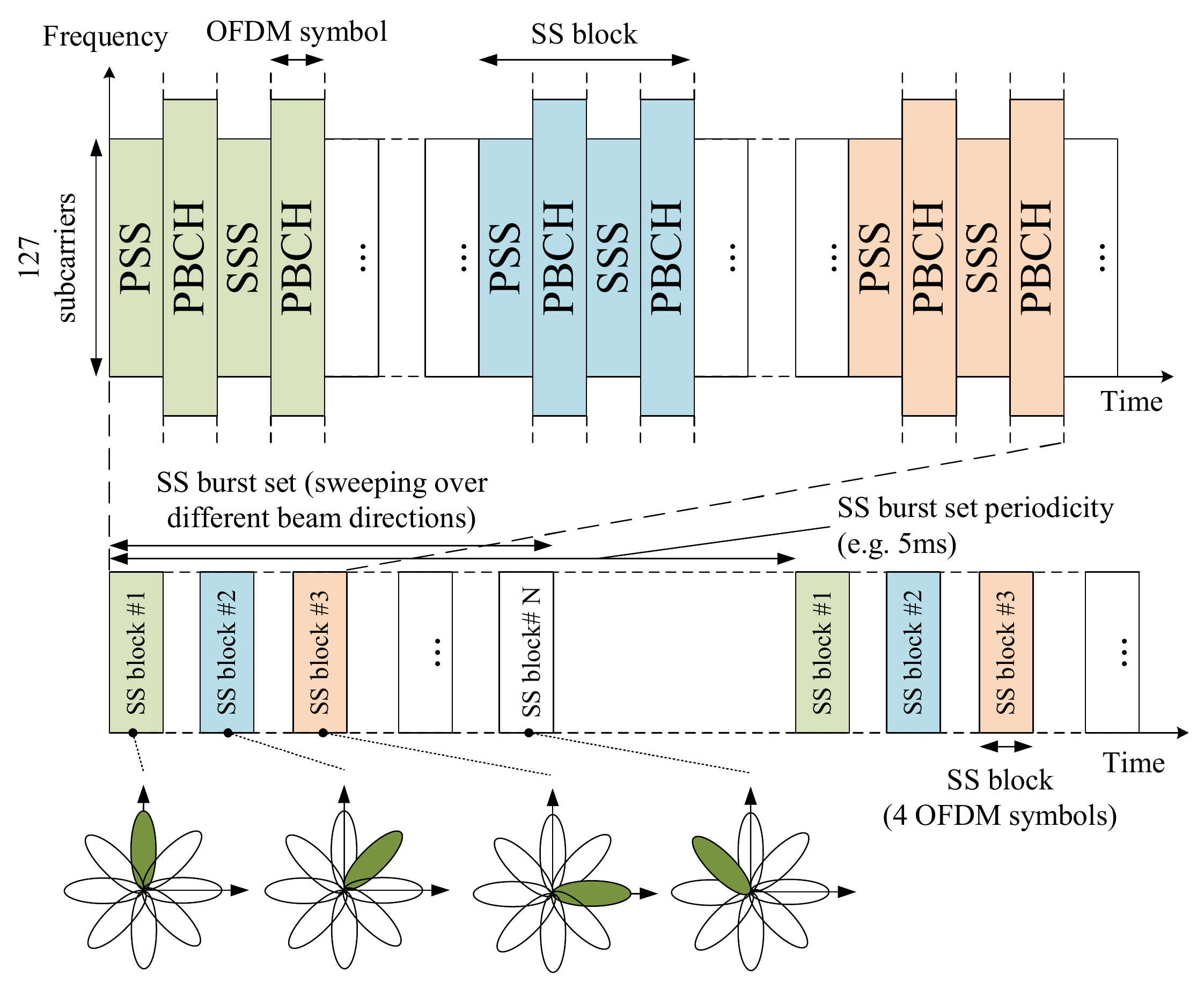}
    \caption{Illustration of the structure of SS block and SS burst set}
    \label{fig:SS_block}
\end{figure}

Synchronization signals are transmitted in form of SS-blocks, as specified in \cite{TS_38211_signal}. The SS-block, illustrated in Fig. \ref{fig:SS_block}, includes a PSS, SSS and PBCH signal. Each SS-block reserves 24 consecutive PRBs in frequency domain and four OFDM symbols in time domain. The structure of PSS and SSS is based on m-sequences, which are mapped within 6 center PRBs in the middle of the SS-block allocation in 127 consecutive subcarriers. Depending on the given physical identity of the base station, there are 3 different PSS and 336 different SSS sequences available. 

In 5G NR, also the synchronization signals are transmitted by using beamforming methods. Thus, in order to cover user terminals in different locations, SS blocks are transmitted in separate directions. Now, a set of SS blocks, which covers one cycle of SS-block transmissions in chosen directions, is called a SS burst set. In this paper, we assume that there are in total 64 SS blocks (i.e. 64 different beam directions) in one SS burst set, and that each time slot from the beginning of the first subframe includes two SS blocks, as described in \cite{SS_block_mapping}, until all SS blocks are transmitted. Thus, the overall duration of transmitting one SS burst set is 2~ms.

\section{Estimation of TOA and AOD based on 5G Synchronization Signals}
\label{sec:TOA_AOD_estimation}

The estimation of TOA and AOD is based on the correlation between the received signal and the known PSS and SSS from separate RRHs. In the considered estimation approach, we assume that the transmitter and receiver clocks are synchronized, and that the time of transmission is known by the receiver. By using (\ref{eq:received_signal}) and considering the distances between the train and the RRHs, the $n^{\text{th}}$ sample of the received signal can be written as
\begin{equation} \label{eq:received_samples}
    \tilde{z}[n] = \sum_{k=0}^{N_{\text{RRH}}-1} \tilde{z}_{k}[n-\delta_{k}] + v[n],
\end{equation}
where $\tilde{z}_{k}[n-\delta_{k}]$ is the sampled signal of $\tilde{z}(t)$ from the $k^{\text{th}}$ RRH, and $\delta_{k}$ is the corresponding propagation delay in samples. In the following denotations, we omit the RRH indices and present the AOD and TOA estimation procedure for an arbitrary RRH. Nonetheless, in order to obtain the AOD and TOA estimates for multiple RRHs, the same process is simply repeated for each available RRH. 

The absolute value of the cross-correlation function over one SS burst set period, between the PSS and SSS sequences from a single RRH, and the received signal $\tilde{z}[n]$, can be given as
\begin{equation} \label{eq:correlation_function}
	r[m] = \left| \sum_{n=0}^{N_{\text{SB}}-1} \tilde{z}^{\ast}[n] \tilde{b}[n+m] \right|,
\end{equation}
where $N_{\text{SB}}$ is the length of the SS burst set in samples, and $\tilde{b}[k]$ is the known reference signal consisting of the PSS and SSS sequences in the dedicated subcarriers with null subcarriers elsewhere. Since there are multiple SS blocks transmitted within one SS burst set, the correlation function includes multiple peaks. The relative amplitude between separate peaks depends on the received signal model given in (\ref{eq:received_power}) and (\ref{eq:received_signal}). However, because the position of the receiver is rather stationary during the reception of one SS burst set, the most significant effect on the relative correlation peak values is the AOD, which is altered between each SS block. Of course, also the Doppler phenomenon affects the temporal signal strengths, but this is expected to change the power levels gradually over adjacent correlation peaks. 

By assuming the knowledge of the used AOD for each SS block in the SS burst set, the AOD estimation can be based on the observation of relative amplitude levels of the measured correlation peaks. Due to the channel, noise and interference, some of the SS blocks do not have a distinguishable correlation peak. Based on the known SS block locations in the signal, we are able to divide the correlation function in (\ref{eq:correlation_function}) into separate SS-block-wise correlation functions as
\begin{equation} \label{eq:SS_wise_correlatiion}
\begin{gathered}
	\tilde{r}_{i}[k] = r[\upsilon_{i} + k] \text{ for } k=0,....,K-1 \\
    \text{ with } K = \min_{i,j \in \Omega_{\text{peak}}} \lvert \upsilon_{i}-\upsilon_{j} \rvert,
\end{gathered}
\end{equation}
where $\tilde{r}_{i}[k]$ and $\upsilon_{i}$ are the correlation function and the first sample index for the $i^{\text{th}}$ SS block, and $\Omega_{\text{peak}}$ is a set of all detected SS block indices. Moreover, the $i^{\text{th}}$ SS block is considered to be successfully detected, if 
\begin{equation}
	\eta_{i} = \max_{k} \tilde{r}_{i}[k] > \mu_{i} + 4\sigma_{i}, 
\end{equation}
where $\eta_{i}$, $\mu_{i}$ and $\sigma_{i}$ are the maximum value, the mean value and the standard deviation of the correlation function of the $i^{\text{th}}$ SS block.

In order to average out noise and to enable AOD estimates to be obtained between the discrete AOD angles used in the SS-block-wise beamforming, the AOD estimate $\hat{\theta}_{\text{T}}$ for the train position under this SS burst set period is obtained by taking a weighted average of three largest correlation peaks as
\begin{equation}
\begin{gathered}
	\hat{\theta}_{\text{T}} = w_{j} \theta_{\text{T},j} \text{ with} \\
    w_{j} =\frac{\eta_{j}}{\sum_{l \in \Omega_{\text{largest}}}\eta_{l}} \text{ and } j \in \Omega_{\text{largest}} \text{ where} \\
    \eta_{j} = \max_{k} \tilde{r}_{j}[k]
\end{gathered}
\end{equation}
where $\theta_{\text{T},j}$ is the used AOD angle of the $j^{\text{th}}$ SS block, $\Omega_{\text{largest}}$ is the set of three SS block indices with the largest correlation values, and $\tilde{r}_{j}[k]$ is the SS-block-wise correlation function given in (\ref{eq:SS_wise_correlatiion}).

When considering the TOA estimation with the assumed LOS conditions, we desire to exploit the cyclic nature of the observed correlation function over all detected SS blocks. Therefore, in order to include timing information from multiple SS blocks, we combine the correlation functions of separate SS blocks as
\begin{equation}
	\tilde{r}_{\text{comb}}[k]=\sum_{i \in \Omega_{\text{peak}}} \tilde{r}_{i}[k].
\end{equation}
After this, the TOA estimate in sample durations can be determined as
\begin{equation}
	\hat{\delta} = \argmax_{k} \tilde{r}_{\text{comb}}[k],
\end{equation}
and the corresponding TOA estimate (in seconds) as $\hat{\tau}=\hat{\delta}/F_{\text{s}}$.

\section{Tracking of the Train}
\label{sec:position_estimation}

The considered tracking approach is based on the EKF \cite{optimal_filter_EKF}, which is able to handle non-linear state-transition and measurement models via function linearization around the currently available estimate of the train state. Besides tracking the position of the train, we also consider tracking of velocity and acceleration. Consequently, the state vector of the train at a time instant $n$ is determined as
\begin{equation}
	\mathbf{s}[n] = \left[x[n],y[n],v_{x}[n],v_{y}[n],a_{x}[n],a_{y}[n]\right]^{\text{T}},
\end{equation}
where $x[n]$, $y[n]$, $v_{x}[n]$, $v_{y}[n]$, $a_{x}[n]$ and $a_{y}[n]$ are the x-coordinate and  the y-coordinate, the velocity in the x-axis and the y-axis, and the acceleration in the x-axis and the y-axis, respectively. We assume a system with a linear state-transition model and a non-linear measurement model given as
\begin{align} \label{eq:EKF_steps}
\begin{split}
    \mathbf{s}[n] &= \mathbf{F}\mathbf{s}[n-1] + \mathbf{q}[n] \\
    \mathbf{y}[n] &= \mathbf{h}\left(\mathbf{s}[n]\right) + \mathbf{w}[n],
\end{split}
\end{align}
where $\mathbf{F}$ is the state-transition matrix, $\mathbf{h}\left(\mathbf{s}[n]\right)$ is the non-linear measurement function as a function the state $\mathbf{s}[n]$, and $\mathbf{q}[n] \sim \mathcal{N}(0,\mathbf{Q})$ and $\mathbf{w}[n] \sim \mathcal{N}(0,\mathbf{W})$ are the process noise and measurement noise vectors. At each estimation time step, the tracking process includes two separate phases, the prediction phase and update phase. In the prediction phase, the a priori estimate of the state $\hat{\mathbf{s}}^{-}[n]$ can be obtained by using the a posteriori estimate of the state $\hat{\mathbf{s}}^{+}[n-1]$ as
\begin{align}
\begin{split}
    \hat{\mathbf{s}}^{-}[n] &= \mathbf{F}\hat{\mathbf{s}}^{+}[n-1]  \\
    \hat{\mathbf{P}}^{-}[n] &= \mathbf{F}\hat{\mathbf{P}}^{+}[n-1]\mathbf{F}^{\text{T}} + \mathbf{Q}
\end{split}
\end{align}
where $\hat{\mathbf{P}}^{-}[n]$ and $\hat{\mathbf{P}}^{+}[n]$ are the a priori and a posteriori estimates of the state covariance. After this, based on the available measurements $\mathbf{y}[n]$, the a posteriori estimates in the update phase can be obtained as
\begin{align} \label{eq:update_phase}
\begin{split}
	\mathbf{K}[n] &= \hat{\mathbf{P}}^{-}[n] \mathbf{H}[n]^{\text{T}}\left( \mathbf{H}[n] \hat{\mathbf{P}}^{-}[n] \mathbf{H}[n]^{\text{T}} + \mathbf{W} \right)^{-1} \\
    \hat{\mathbf{s}}^{+}[n] &= \hat{\mathbf{s}}^{-}[n] + \mathbf{K}[n] \left( \mathbf{y}[n]-\mathbf{h}\left(\hat{\mathbf{s}}^{-}[n]\right) \right) \\
    \hat{\mathbf{P}}^{+}[n] &= \left(\mathbf{I} - \mathbf{K}[n]\mathbf{H}[n] \right) \hat{\mathbf{P}}^{-}[n]
\end{split}
\end{align}
where $\mathbf{K}[n]$ denotes the Kalman gain, and $\mathbf{H}[n]$ is the Jacobian matrix of the measurement function $\mathbf{h}(\cdot)$ evaluated at $\hat{\mathbf{s}}^{-}[n]$. 

We consider the acceleration of the train to be nearly constant between two consecutive time instants of the state. Therefore, we define the linear state-transition matrix based on the continuous acceleration model, used in \cite{5G_mike_cars}, as
\begin{equation}
	\mathbf{F} = \left[ \begin{matrix} \mathbf{I}_{2\times2} & \Delta t \mathbf{I}_{2\times2} & \frac{\Delta t^{2}}{2} \mathbf{I}_{2\times2} \\ \mathbf{0}_{2\times2} & \mathbf{I}_{2\times2} & \Delta t \mathbf{I}_{2\times2} \\ \mathbf{0}_{2\times2} & \mathbf{0}_{2\times2} & \mathbf{I}_{2\times2} \end{matrix} \right],
\end{equation}
where $\Delta t$ is the time interval between two consecutive states. Furthermore, the corresponding process noise covariance can be given as
\begin{equation}
	\mathbf{Q} = \sigma_{\text{a}}^{2} \left[ \begin{matrix} \frac{\Delta t^{5}}{20} \mathbf{I}_{2\times2} & \frac{\Delta t^{4}}{8} \mathbf{I}_{2\times2} & \frac{\Delta t^{3}}{6} \mathbf{I}_{2\times2} \\ \frac{\Delta t^{4}}{8} \mathbf{I}_{2\times2} & \frac{\Delta t^{3}}{3} \mathbf{I}_{2\times2} & \frac{\Delta t^{2}}{2} \mathbf{I}_{2\times2} \\ \frac{\Delta t^{3}}{6}\mathbf{I}_{2\times2} & \frac{\Delta t^{2}}{2} \mathbf{I}_{2\times2} & \Delta t \mathbf{I}_{2\times2} \end{matrix} \right],
\end{equation}
where $\sigma_{\text{a}}^{2}$ denotes the variance of the acceleration noise.

By considering the AOD and TOA estimates as given in Section \ref{sec:TOA_AOD_estimation}, the measurement vector can be written as $\mathbf{y}[n]=[\mathbf{y}_{0}[n],...,\mathbf{y}_{\hat{N}_{\text{R}}-1}[n]]^{\text{T}}$, where $\mathbf{y}_{k}[n] = [\hat{\theta}_{\text{T},k},\hat{\tau}_{k}]^{\text{T}}$ includes the AOD and TOA estimates from a single RRH, and $\hat{N}_{\text{R}-1}$ is the total number of available RRHs. Now, for the $k^{\text{th}}$ available RRH, the non-linear measurement function, related to the measurement vector $\mathbf{y}_{k}[n]$, can be written as
\begin{equation} \label{eq:measurement_function}
	\mathbf{h}_{k}\left(\mathbf{s}[n]\right) = \left[\begin{matrix} 
    \arctan{\frac{\Delta y_{k}[n]}{\Delta x_{k}[n]}} \\
    \frac{\lVert \mathbf{p}[n]-\mathbf{p}_{k} \rVert}{c}
    \end{matrix}\right],
\end{equation}
where $\Delta y_{k}[n] = y[n]-y_{k}$ and $\Delta x_{k}[n] = x[n]-x_{k}$ are the distances between the $k^{\text{th}}$ available RRH and the train in the x direction and y direction, $\mathbf{p}_{k} = [x_{k}, y_{k}]^{\text{T}}$ and $\mathbf{p}[n] = [x[n], y[n]]^{\text{T}}$ are the position of the $k^{\text{th}}$ available RRH and the train, and $c$ is the speed of light. Moreover the overall measurement function, including all the available RRHs, can be written as $\mathbf{h}\left(\mathbf{s}[n]\right) = [\mathbf{h}_{0}\left(\mathbf{s}[n]\right),...,\mathbf{h}_{\hat{N}_{\text{R}}-1}\left(\mathbf{s}[n]\right)]^{\text{T}}$. Finally, the Jacobian matrix $\mathbf{H}[n] \in \mathbb{R}^{2\hat{N}_{\text{R}} \times 6}$ in (\ref{eq:update_phase}) can be written as 
\begin{equation} \label{eq:jacobian_matrix}
	\mathbf{H}[n] = \left[ \begin{matrix} 
    \frac{-\Delta \hat{y}_{0}[n]}{\lVert \hat{\mathbf{p}}[n]-\mathbf{p}_{0} \rVert^{2}} &
    \frac{\Delta \hat{x}_{0}[n]}{\lVert \hat{\mathbf{p}}[n]-\mathbf{p}_{0} \rVert^{2}} & \mathbf{0}_{1 \times 4} \\
    \frac{\Delta \hat{x}_{0}[n]}{c \lVert \hat{\mathbf{p}}[n]-\mathbf{p}_{0} \rVert} &
    \frac{\Delta \hat{y}_{0}[n]}{c \lVert \hat{\mathbf{p}}[n]-\mathbf{p}_{0} \rVert} & \mathbf{0}_{1 \times 4} \\
    \vdots & \vdots & \vdots \\
    \frac{-\Delta \hat{y}_{\hat{N}_{\text{R}}-1}[n]}{\lVert \hat{\mathbf{p}}[n]-\mathbf{p}_{\hat{N}_{\text{R}}-1} \rVert^{2}} &
    \frac{\Delta \hat{x}_{\hat{N}_{\text{R}}-1}[n]}{\lVert \hat{\mathbf{p}}[n]-\mathbf{p}_{\hat{N}_{\text{R}}-1} \rVert^{2}} & \mathbf{0}_{1 \times 4} \\
    \frac{\Delta \hat{x}_{\hat{N}_{\text{R}}-1}[n]}{c \lVert \hat{\mathbf{p}}[n]-\mathbf{p}_{\hat{N}_{\text{R}}-1} \rVert} & 
    \frac{\Delta \hat{y}_{\hat{N}_{\text{R}}-1}[n]}{c \lVert \hat{\mathbf{p}}[n]-\mathbf{p}_{\hat{N}_{\text{R}}-1} \rVert} & \mathbf{0}_{1 \times 4}
    \end{matrix}\right]
\end{equation}
where the elements are obtained by, first, taking partial derivatives of the non-linear measurement function in (\ref{eq:measurement_function}) with respect to the state vector $\mathbf{s}[n]$, and second, by evaluating the resulted Jacobian matrix at the estimated a priori state $\hat{\mathbf{s}}^{-}[n]$.

\section{Positioning Results and Analysis}
\label{sec:results}

\begin{figure}[t]
    \centering
    \includegraphics[width=\columnwidth]{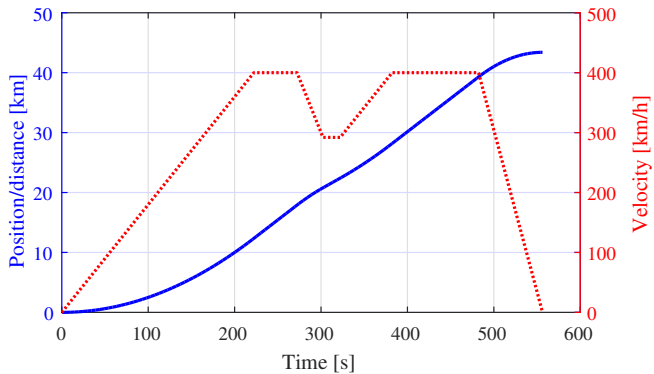}
    \caption{The position (solid line) and velocity (dotted line) of the train in the x direction during the simulation. The y-coordinate is fixed to zero.}
    \label{fig:track}
\end{figure}

In order to study the SS-block-based positioning of a high speed train, we consider a track of over 43 km with variable train velocity of up to 400 km/h. This train movement model is rather similar to the one used in \cite{locationAware5GForHST}, where location-aware 5G communications were studied in the HST scenario. Due to the assumed high-speed scenario, the maximum curvature of the railroad is very limited. Based on the HST scenario specified in \cite{TR_38913_scenario}, and the related system geometry shown in Fig. \ref{fig:system}, the track can be approximated as a straight line in the vicinity of the closest RRHs from which the positioning signals are obtained. Consequently, we fix the y-coordinate of the railroad to zero, and thus, the train position and train velocity are fully determined by the x coordinate and its derivative. The profile of the track is shown in Fig. \ref{fig:track}, where both the position and velocity in the x-direction are illustrated as a function of time.

\begin{figure}[b]
    \centering
    \includegraphics[width=\columnwidth]{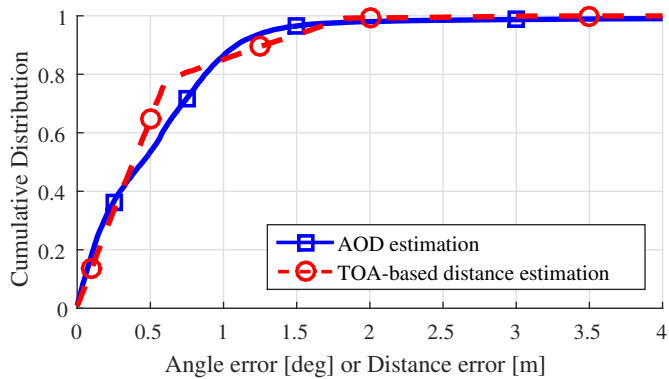}
    \caption{Cumulative error distribution of the AOD estimates (Angle error) and TOA-based distance estimates (Distance error).}
    \label{fig:TOA_DOA_cumerror}
\end{figure}

As shown in Fig. \ref{fig:system}, the RRHs are located uniformly on the upper side of the railroad with 500 m intervals and with 15~m distance to the railroad. Moreover, the RRHs are synchronized to sweep the SS burst set of 64 separately beamformed SS blocks uniformly from 0~deg to -180~deg angles. The used carrier frequency is 30~GHz and the transmission power of each RRH is fixed to 33~dBm. Furthermore, each RRH has two antenna panels mounted towards -45~deg and -135~deg directions. The RRHs are given separate PSS/SSS sequence identities, so that the same combination of PSS and SSS sequences cannot be heard in the same location from multiple RRHs, and the PSS sequences of adjacent RRHs are always different. Besides the SS-blocks, each RRH transmits user data over the whole band with full buffer traffic model.

We assume that the train obtains AOD and TOA measurements from the SS burst sets in intervals of 100~ms, by using the estimators described in Section \ref{sec:position_estimation}. Depending on the SS burst set periodicity, measurements could also be taken with considerably higher interval, such as 5 ms, which would improve the tolerance against noise and interference. For simplicity, AOD and TOA measurements are taken into account from only at most three highest power RRHs. We use the channel model described in Section \ref{sec:system_model}, and we define the noise figure of the train receiver as 5~dB and the fundamental thermal noise power density as -174~dBm/Hz. However, it should be noticed that the properties of the radio channel, more specifically the Doppler spread, vary during the simulation along with the train velocity. In the train, there are two antenna panels mounted towards the nose and tail of the train, and the antenna beams are fixed towards the panel directions.

The cumulative distributions of the estimation error of AOD and the TOA-based distance are shown in Fig. \ref{fig:TOA_DOA_cumerror}, where the values of the x-axis are defined in degrees for the AOD-estimation-based angle error, and in meters for the TOA-estimation-based distance error. It can be observed that 95~\% of the estimation error values are below 1.3 deg for the AOD estimates, and below 1.6 m for the TOA-based distance estimates. Due to the discretized nature of the TOA estimation, where the resolution of the propagation delay estimate is limited to the used sample rate, the cumulative error distribution function of the TOA-based distance estimation has distinctive linearly behaving error regions. Because the used sample rate is $F_{\text{s}}=245.76 \text{ MHz}$, the resolution of distance estimation is $c/F_{\text{s}} \approx \text{1.21~m}$. With an optimal sample delay estimate, this results in $\pm\text{0.6~m}$ error region, which can be observed in the TOA-based distance estimation curve in Fig. \ref{fig:TOA_DOA_cumerror}.

\begin{figure}[t]
    \centering
    \includegraphics[width=\columnwidth]{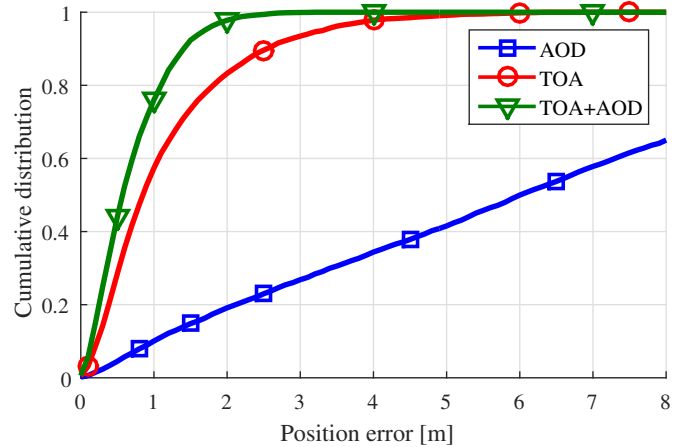}
    \caption{Cumulative error distribution of the train position error for considering, only AOD measurements, only TOA measurements, and the both measurements (AOD+TOA)}
    \label{fig:pos_error_all}
\end{figure}

The cumulative error distribution for the absolute positioning error is shown in Fig. \ref{fig:pos_error_all} for using only AOD measurement, only TOA measurements, and both measurements (AOD +TOA). Due to the specific system geometry, where the AOD is nearly constant for a long time when the train is between the RRHs, using only AOD does not provide satisfactory positioning performance. However, since the TOA-based estimation has a degraded performance only in proximity to a measured RRH, and most of the time the train is relatively far away from measured RRHs, the TOA-based approach is able to provide significantly improved positioning performance. Nonetheless, when using jointly both the AOD and TOA measurements, the positioning performance can be yet improved, as the two measurement types complement each other from the system geometry perspective. From the tracking point of view, the difference in computational load between using only AOD measurements, only TOA measurements, or jointly both of the measurements, is relatively small, since the set of used measurements affect only the number of rows of the Jacobian matrix given in (\ref{eq:jacobian_matrix}). By considering both the AOD and TOA measurement, the mean error accuracy is 0.66~m, which is below the 1m 5G target, and moreover, 95\% of the positioning errors are below 1.7~m, and 99\% below 2.3~m. The targeted submeter positioning accuracy can be achieved in approximately 75~\% of the time. In addition, although not seen in the figure, it was noticed that the positioning accuracy was similar throughout separate values of train acceleration. However, whenever the train acceleration changes, the tracking algorithm introduces a lag, which affects the tracking accuracy over a certain state convergence period.

\section{Conclusion and Discussion}
\label{sec:conclusions}

In this paper, we studied positioning performance of a high-speed train by using 5G NR mobile network and its specific synchronization signals. We first developed practical TOA and AOD estimators building on the processing of received synchronization signals. By using the channel models described for the considered scenario, we then simulated a train track of length more than 43 km with variable train velocity of up to 400 km/h. Based on AOD and TOA measurements and the EKF-based tracking model, we obtained a mean positioning accuracy of 0.66 m with 95\% errors below 1.7m. Due to the specific system geometry, significantly better positioning performance was obtained by using only TOA measurements compared to using only AOD measurements. However, by including both the TOA and AOD measurements in the used EKF tracking algorithm, the positioning performance was further improved and achieved a mean tracking accuracy below the 1~m 5G target value. For the future studies, it is important to consider different positioning environments with appropriate channel models. Furthermore, including the effect of clock synchronization errors on the tracking solutions and positioning performance evaluations, would provide important insight into the practical feasibility of the solution. However, since the presented positioning performance is already at a very encouraging level, it can be argued that there is substantial potential in the forthcoming 5G-based positioning solutions for high-speed trains.

The achieved submeter accuracy with 75~\% availability is sufficient for many non-critical use cases, such as displaying the train position for the passengers. However, with mission-critical use cases, such as autonomous trains, the 75~\% availability is not sufficient, but fortunately, the real time positioning accuracy can be monitored by the proposed EKF via the included state covariance matrices (i.e. $\hat{\mathbf{P}}^{-}[n]$ and $\hat{\mathbf{P}}^{+}[n]$). Depending on the use-case-specific availability requirements, the positioning performance can be upgraded by fusing the proposed 5G positioning approach with other types of positioning technologies available in the train systems, such as GNSS, odometer and Doppler radar. Consequently, with appropriate data fusion, the overall positioning performance has great potential to achieve the requirements for numerous mission-critical use cases, such as the envisioned autonomous trains.



\ifCLASSOPTIONcaptionsoff
  \newpage
\fi

\bibliographystyle{IEEEbib}
\bibliography{refs}

\end{document}